\documentclass[preprint,showpacs,preprintnumbers,amsmath,amssymb]{revtex4}

\usepackage{graphicx}
\usepackage{bm}

\begin{document}

\title{Effective Action for a Statistical System with a Field dependent Wave
Function}
\author{Pierre Gosselin$^{a}$, Herv\'{e} Mohrbach$^{b}$, Alain B\'{e}rard$^{c}$} 

\affiliation{ a) Universit\'{e} Grenoble I, Institut Fourier, UMR\ 5582 CNRS-UJF, UFR de
Math\'{e}matiques,
BP74, 38402 Saint Martin d'H\`{e}res, Cedex, France} 
\affiliation{ b) Institut Charles Sadron, CNRS UPR 022, 6 rue Boussingault, 
67083 Strasbourg Cedex, France}
\affiliation{ c) L.P.L.I. Institut\ de\ Physique, 1 blvd D.Arago, F-57070, Metz, France.}

\date{\today}

\begin{abstract}
We compute the first order correction in $\hbar $ to the field dependent
wave function in Statistical Field Theory. These corrections are evaluated
by several usual methods. We limit ourselves to a one dimensional model in
order to avoid the usual difficulties with the UV divergences that are not
relevant for our purposes. The main result of the paper is that the various
methods yield different corrections to the wave function. Moreover, we give
arguments to show that the perturbative integration of the exact
renormalization group provides the right result.
\end{abstract}

\maketitle

\section{Introduction}

Recently the first order quantum correction in $\hbar $ (one-loop) to the
classical action of a quantum particle in one dimension, has been computed
by Cametti et al.\cite{cametti}. These corrections evaluated to the second
order in the derivative expansion of the effective action, using the
Iliopoulos et al.\cite{iliopoulos} expansion, lead to a renormalization of
both the potential and the kinetic energy. Borelli and Kleinert\cite{borelli}
have extended this calculation to a theory with a field dependent wave
function. Instead of using the expansion of Iliopoulos et al.\cite
{iliopoulos}, they used a method of Frazer\cite{frazer} which can be
generalized to terms of higher order in derivatives. Note that usually a
quantum mechanical system can be considered as a statistical system in one
dimension. However, this identification is no more correct for an
Hamiltonian with a position dependent mass term. Actually, in quantum
mechanics one has to ask for an additional reparametrization invariance\cite
{kl2}. In this paper we will consider our system as a one dimensional model
in statistical physics. To extrapolate our results in quantum mechanics,
adding the one loop contribution due to the reparametrization invariance
would be necessary.

\bigskip

In this paper we aim to compare several approaches for the computation of
the one loop wave function renormalization. In order to do so, it is
instructive enough to consider a $1D$ theory in order to avoid the usual UV
complications. In particular, we first compute the one loop-renormalization
of \ the kinetic energy with a field dependent wave functions by using a
more general expansion than the Frazer's one. This method introduced by Zuk 
\cite{zuk} does not need an expansion about a constant field and leads to
different corrections from the ones obtained in \cite{borelli}. Our result
is a different expression for the effective wave function than the
Borelli-Kleinert one.

Two other methods based on the perturbative integration of the 'exact' RG
equation are also considered. The first one is a RG flow obtained by the
mean of a regulator -similar to a smooth cutoff- in the Schwinger Proper
Time formalism\cite{Zappa}. Surprisingly it leads to the same result as the
one obtained with the method of Zuk. However the weakness of this RG
equation is the lack of first principle in its derivation.

The second RG approach is based on another exact equation (ERG) derived in 
\cite{Pierre}. It's advantage in comparison to the preceding approach relies
on an exact Fourier modes after modes integration in the path integral. This
integration is performed without any device like a smooth cutoff. The ERG
equation thus obtained as well as the resulting expression for the wave
function renormalization are different from the other ones. We provide an
explanation for this fact : the perturbative expansion and the smooth cutoff
RG method rely on an integration over the all Fourier space, including paths
far from the classical one, leading to additional wrong contributions. As a
consequence the ERG equation seems to provide the correct one loop wave
function renormalization.

\section{Effective Action}

The effective action formalism in stastistical physics is explained in \cite
{cametti}. Here we follow the approach of paper\cite{borelli}.

The starting point is the action in one dimension with a field dependent
wave function: 
\begin{equation}
\mathcal{A}\left[ \varphi \right] =\int\nolimits_{t_{a}}^{t_{b}}\left( \frac{%
Z\left( \varphi \right) }{2}\overset{.}{\varphi }^{2}-V\left( \varphi
\right) \right) dt  \label{action}
\end{equation}
The partition function \cite{kleinertbook} reads in the semi-classical
approximation \cite{borelli}: 
\begin{equation}
Z_{QM}\left( t_{b,}t_{a}\right) \left[ \varphi _{cl}\right] =e^{\frac{i}{%
\hbar }\mathcal{A}\left[ \varphi _{cl}\right] }\int\limits_{\varphi
_{a}=\varphi _{b}}\mathcal{D}\overline{\varphi }e^{\frac{i}{2\hbar }%
\int\nolimits_{t_{a}}^{t_{b}}dt\overline{\varphi }\left( t\right) K\left(
t\right) \overline{\varphi }\left( t\right) }  \label{partition}
\end{equation}
where $\overline{\varphi }\left( t\right) =\varphi \left( t\right) -\varphi
_{cl}\left( t\right) $ are the fluctuations around the classical path
solution of the classical equation of motion, and $K\left( t\right) =\frac{%
\delta ^{2}\mathcal{A}}{\delta \varphi \left( t\right) \delta \varphi \left(
t\right) }\mid _{\varphi cl}$. The semi-classical effective action is
defined as: 
\begin{equation*}
\mathcal{A}_{eff}\left[ \varphi _{cl}\right] =-i\hbar \ln Z_{QM}\left[
\varphi _{cl}\right]
\end{equation*}
which after the gaussian integration in (\ref{partition}) reads: 
\begin{equation}
\mathcal{A}_{eff}\left[ \varphi _{cl}\right] =\mathcal{A}\left[ \varphi _{cl}%
\right] -\frac{i\hbar }{2}\text{Tr}\ln K\left( t\right)  \label{effective}
\end{equation}
with: 
\begin{equation*}
K\left( t\right) =Z\left( \varphi _{cl}\right) \widehat{\omega }%
^{2}-V^{\prime \prime }\left( \varphi _{cl}\right) -iZ^{\prime }\left(
\varphi _{cl}\right) \overset{.}{\varphi }_{cl}\widehat{\omega }+\frac{1}{2}%
Z"\left( \varphi _{cl}\right) \overset{.}{\varphi }_{cl}^{2}-\frac{d}{dt}%
\left( Z^{\prime }\left( \varphi _{cl}\right) \overset{.}{\varphi }%
_{cl}\right)
\end{equation*}
In \cite{borelli} the one-loop correction was computed by setting $\varphi
_{cl}\left( t\right) $ equal to $\varphi _{0}+\widetilde{\varphi }\left(
t\right) $, where $\varphi _{0}$ is constant and expanding the logarithmic
term in (\ref{effective}) in powers of $\ \widetilde{\varphi }\left(
t\right) $ and its derivatives.

\bigskip

The method of Zuk\cite{zuk} we use here, is based on a different kind of
expansion in which it is not necessary to expand the classical path $\varphi
_{cl}\left( t\right) $ around a constant path. The expansion is obtained by
defining $K\left( u\right) =K\left( t\right) +u$, where $u$ is a $($mass$%
)^{2}$ parameter, then by deriving with respect to this parameter and
expanding the logarithmic term in (\ref{effective}). In fact by defining $%
\Gamma \left( 0\right) =\frac{i\hbar }{2}Tr\ln K\left( t\right) $ we have
the following relation: 
\begin{equation}
\Gamma \left( 0\right) =-\int\nolimits_{0}^{\infty }du\frac{d}{du}\Gamma
\left( u\right) =-\frac{1}{2}\int\nolimits_{0}^{\infty }du\text{Tr}%
K^{-1}\left( u\right)  \label{gamm}
\end{equation}
Now we follow Zuk\cite{zuk} by writing the expansion: 
\begin{equation*}
K^{-1}\left( u\right) =A^{-1}-A^{-1}BA^{-1}+A^{-1}BA^{-1}BA^{-1}+...
\end{equation*}
where $A=Z(\varphi _{cl})w^{2}-V^{^{\prime \prime }}\left( \varphi
_{cl}\right) -u$ and $B$ contains all the other contributions of the
operator $K(t).$ Note that we have introduced the eigenvalue $w^{2}$ of the
energy operator $\widehat{w}$ whose coordinate time representation is $%
-id/dt $. Following the same steps than in \cite{zuk} we collect in (\ref
{gamm}) all the contributions to the coefficient of \ $\overset{.}{\varphi }%
_{cl}^{2} $in order to get the effective wave function. A lengthy
computation leads to: 
\begin{eqnarray}
Z_{eff}\left( \varphi _{cl}\right) &=&Z\left( \varphi _{cl}\right) +\frac{%
\hbar }{32}\frac{\left[ V^{\prime \prime \prime }\left( \varphi _{cl}\right) %
\right] ^{2}\left[ Z\left( \varphi _{cl}\right) \right] ^{1/2}}{\left[
V^{\prime \prime }\left( \varphi _{cl}\right) \right] ^{5/2}}-\frac{3\hbar }{%
16}\frac{V^{\prime \prime \prime }\left( \varphi _{cl}\right) Z^{\prime
}\left( \varphi _{cl}\right) }{\left[ Z\left( \varphi _{cl}\right) \right]
^{1/2}\left[ V^{\prime \prime }\left( \varphi _{cl}\right) \right] ^{3/2}} 
\notag \\
&&-\frac{7\hbar }{32}\frac{\left[ Z^{\prime }\left( \varphi _{cl}\right) %
\right] ^{2}}{\left[ Z\left( \varphi _{cl}\right) \right] ^{3/2}\left[
V^{\prime \prime }\left( \varphi _{cl}\right) \right] ^{1/2}}+\frac{\hbar }{4%
}\frac{Z^{\prime \prime }\left( \varphi _{cl}\right) }{\left[ Z\left(
\varphi _{cl}\right) \right] ^{1/2}\left[ V^{\prime \prime }\left( \varphi
_{cl}\right) \right] ^{1/2}}  \label{meff}
\end{eqnarray}

The important and intriguing result is that the various terms in the
one-loop expression of $Z_{eff}\left( \varphi _{cl}\right) $ have the same
form as those obtained in \cite{borelli}, but the numerical coefficients are
different. For a field-independent wave function we recover the same result
found in \cite{borelli} and \cite{cametti}.

\section{One loop computation from Schwinger Proper Time Formalism.}

It is well known that the effective action can be computed
non-perturbatively through the Renormalization Group (RG) method. Recently a
particular version of the RG flow obtained by means of regulator in the
Schwinger Proper Time (PT) formalism has been introduced \cite{Zappa}. This
method was used to compute the energy gap between the first excited state
and the ground state energy of a one quantum particle system. This work was
an alternative to previous computation we performed by means of the exact RG
for various quantum systems \cite{Pierre}.

The flow equation for the running $Z_{k}$ in the PTRG formalism is given in 
\cite{Zappa}: 
\begin{eqnarray}
k\frac{\partial Z_{k}}{\partial k} &=&\left( \frac{k^{2}}{4\pi }\right) ^{%
\frac{1}{2}}e^{-V^{\prime \prime }/Z_{k}k^{2}}  \notag \\
&&\times \left( -\frac{Z_{k}^{\prime \prime }}{Z_{k}k^{2}}+\frac{21\left(
Z_{k}^{\prime }\right) ^{2}}{24Z_{k}^{2}k^{2}}+\frac{9Z_{k}^{\prime
}V^{\prime \prime \prime }}{6\left( Z_{k}k^{2}\right) ^{2}}-\frac{%
Z_{k}\left( V^{\prime \prime \prime }\right) ^{2}}{6\left( Z_{k}k^{2}\right)
^{3}}\right)
\end{eqnarray}
To compute the one-loop effective wave function we integrate the previous
equation from $0$ to $\infty $ keeping the kinetic energy and the potential
to their bare values. We directly obtain for the effective wave function: 
\begin{eqnarray}
Z_{eff}\left( \varphi _{cl}\right) &=&Z\left( \varphi _{cl}\right) +\frac{%
\hbar }{32}\frac{\left[ V^{\prime \prime \prime }\left( \varphi _{cl}\right) %
\right] ^{2}\left[ Z\left( \varphi _{cl}\right) \right] ^{1/2}}{\left[
V^{\prime \prime }\left( \varphi _{cl}\right) \right] ^{5/2}}-\frac{3\hbar }{%
16}\frac{V^{\prime \prime \prime }\left( \varphi _{cl}\right) Z^{\prime
}\left( \varphi _{cl}\right) }{\left[ Z\left( \varphi _{cl}\right) \right]
^{1/2}\left[ V^{\prime \prime }\left( \varphi _{cl}\right) \right] ^{3/2}} 
\notag \\
&&-\frac{7\hbar }{32}\frac{\left[ Z^{\prime }\left( \varphi _{cl}\right) %
\right] ^{2}}{\left[ Z\left( \varphi _{cl}\right) \right] ^{3/2}\left[
V^{\prime \prime }\left( \varphi _{cl}\right) \right] ^{1/2}}+\frac{\hbar }{4%
}\frac{Z^{\prime \prime }\left( \varphi _{cl}\right) }{\left[ Z\left(
\varphi _{cl}\right) \right] ^{1/2}\left[ V^{\prime \prime }\left( \varphi
_{cl}\right) \right] ^{1/2}}
\end{eqnarray}
which is the same result than the one obtained by the one-loop computation
by means of the method of Zuk.

\section{One loop computation from Exact Renormalization Group.}

It is interesting to compare the preceding result with an exact RG equation
approach we derived previously \cite{Pierre}. Working in discrete time
allowed us to compute the RG flow one mode after the other in the path
integral. The main point is that the RG equation for the effective action is
an exact (non-perturbative) equation obtained without any approximation.
Note that this is possible only in one dimension since in higher dimension
the computation is plagued by non-analytical terms as explained in\cite
{Pierre} an below. Note that contrary to our equation, the PTRG equation 
\cite{Zappa} does not have a first principle derivation.

The running wave function equation is\cite{Pierre}: 
\begin{equation}
Z_{n-1}(\varphi _{0})=Z_{n}(\varphi _{0})+{\frac{1}{2\beta }}\left( 1+{\frac{%
{Z}_{n}^{\prime \prime }(\varphi _{0})}{Z\omega _{n}^{2}+V^{\prime \prime }}}%
\right)  \label{rgnous}
\end{equation}
where $n$ denotes the $n$th discrete mode and $\omega _{n}$ is the
corresponding Fourier mode. Remind that $\beta =\frac{\left( N+1\right)
\epsilon }{\hbar }$ is the inverse temperature\cite{kleinertbook} and $%
\epsilon $ is the time interval (lattice spacing) . In the zero temperature
limit $N\epsilon \rightarrow \infty ,$ and the continuum limit $\epsilon
\rightarrow 0$ equation $\left( \text{\ref{rgnous}}\right) $becomes: 
\begin{equation}
\frac{\partial Z_{k}(\varphi _{0})}{\partial k}=\frac{\hbar }{2\pi }\frac{%
Z_{k}^{\prime \prime }(\varphi _{0})}{Z_{k}(\varphi _{0})k^{2}+V_{k}^{\prime
\prime }(\varphi _{0})}  \label{rgrig}
\end{equation}
where $k$ is the energy cut-off. By solving this equation at the one-loop
level we are led to: 
\begin{equation*}
Z_{eff}(\varphi _{0})=Z\left( \varphi _{0}\right) +\frac{\hbar }{4}\frac{%
Z^{\prime \prime }\left( \varphi _{0}\right) }{\left[ Z\left( \varphi
_{0}\right) \right] ^{1/2}\left[ V^{\prime \prime }\left( \varphi
_{0}\right) \right] ^{1/2}}
\end{equation*}
This expression is in contradiction with those obtained by Kleinert and with
the method of Zuk. The only common contribution is the last term in (\ref
{meff}). Note that as a consequence of $\left( \text{\ref{rgrig}}\right) $%
there is no renormalization effect for a constant bare wave function: $%
Z_{eff}(\varphi _{0})=Z\left( \varphi _{0}\right) $. This result agrees with
some numerical computations done in \cite{kroger}.

The absence of the other terms in $\left( \text{\ref{rgrig}}\right) $ has
been already explained in \cite{Pierre} and is related to the presence of
the so-called non-analytical terms when the Fourier modes are continuous. In
field theory these terms appear when the integration on the fast modes is
performed on a continuous spherical shell of thickness $\Delta k$. It is
precisely these terms which give the additional contribution to the flow of
the kinetic energy if we forget the non-analytical terms.

To precise our argument let us consider the derivation of the RG flow for
the wave function by integrating the Fourier modes in a continuous shell of
thickness $\Delta k$. For the point we want to show it is enough to consider
a space independent wave function. We arrive at the following equation (see 
\cite{Pierre}): 
\begin{equation*}
Z_{k-\Delta k}q^{2}+U_{k-\Delta k}^{\left( 2\right) }(\varphi
_{0)}=Z_{k}q^{2}+U_{k}^{\left( 2\right) }(\varphi _{0)}+U_{k}^{\left(
4\right) }(\varphi _{0)}\int_{_{k}}^{_{k-\Delta k}}\frac{dp}{2\pi }\frac{1}{%
G(p)}+F(q)
\end{equation*}
where 
\begin{equation}
F(q)=\frac{\left( U_{k}^{\left( 3\right) }(\varphi _{0)}\right) ^{2}}{4}%
\int_{_{k}}^{_{k-\Delta k}}\frac{dp}{2\pi }\int_{_{k}}^{_{k-\Delta k}}\frac{%
dp^{^{\prime }}}{2\pi }\frac{\delta (p+p^{^{\prime }}+q)+\delta
(p+p^{^{\prime }}-q)}{G(p)G(p^{^{\prime }})}+h.c.  \label{fq}
\end{equation}
It is clear that for $q<\Delta k$, this integral gives a contribution of the
form $\Delta k-\left| q\right| $ because the domain of integration is
deformed by the Dirac delta constraints. By expanding the denominator in
powers of $q$, we obtain the non analytical contribution. : 
\begin{equation*}
F(q)=\frac{\left( U_{k}^{\left( 3\right) }(\varphi _{0)}\right) ^{2}}{4}%
\frac{\Delta k-\left| q\right| }{G^{2}(k)}\left( 1+O(q^{2})\right) \text{
\qquad for }q<\Delta k
\end{equation*}
and 
\begin{equation*}
F(q)=0\text{ \qquad for }q\geqslant \Delta k
\end{equation*}

\smallskip

\smallskip Retrieving the first order quantum correction to the classical
equation derived firstly by Jona-Lasinio's group, as well as
Borelli-Kleinert result, corresponds to neglect the $\left| q\right| $
contribution and let 
\begin{equation*}
F(q)=\frac{\left( U_{k}^{\left( 3\right) }(\varphi _{0)}\right) ^{2}}{4}%
\frac{\Delta k}{G^{2}(k)}
\end{equation*}
However,it is obvious that a rigorous computation of the discrete version of
equation$\left( \text{\ref{fq}}\right) $ leads to the result $F(q)=0$,
yielding formula $\left( \text{\ref{rgrig}}\right) $\smallskip in the limit $%
\Delta k\rightarrow 0$.

The cancellation of the $F(q)$ contribution shows an important difference
between the Renormalization Group and the perturbative expansion. In the
last one, the semi classical contributions are obtained by integrating over
all Fourier space, after Taylor expanding the action around a classical
path. This is problematic, since it corresponds to integrate over all paths
including those far from the classical path. On the contrary, the
Renormalization Group approach, by introducing integration on deformed
shells of possible zero measure (see \cite{Pierre} ), leads to the
cancellation of some terms that are relevant in the perturbative or semi
classical approach, like the term $F(q)$. This cancellation is simply
reminiscent of the difference between integrating a two variables function
over a square $[0,\Lambda ]^{2}$ (as in $\left( \text{\ref{fq}}\right) $,
where the two variables are $p$ and $p^{^{\prime }}$) and the integration
over smalls intervals $[k,k+\Delta k]^{2},k\in \lbrack 0,\Lambda ]$ and then
summing over the squares. In the first case the integral is performed over a
bigger space. Note that this problem only arises for the renormalization of
the wave function and is absent for the computation of the corrections to
the potential energy. Actually in this last case, the perturbative expansion
is based on an integration over $[0,\Lambda ]$ whereas the perturbative
integration of the RG equation introduces a sum of integration over the one
dimensional intervals $[k,k+\Delta k],k\in \lbrack 0,\Lambda ]$. The two
procedures are obviously equivalent in that case.

In that context we can now explain why the perturbative integration of the
PTRG equation leads to the same result as the perturbative method (in
particular the result of Zuk). Indeed, in the derivation of the PTRG
equation one has to replace the double integration of the Dirac delta in $(%
\ref{fq})$ by only one integral with a smooth cutoff. This unique integral
is performed over all the modes, like in the perturbation computation, and
this leads to a non zero contribution of $F(q)$. The smooth cutoff device,
while avoiding the problem of the non-analytical terms, relies on
integration over a too large Fourier space leading to take into account
wrong additional terms.

\smallskip

The difference between the usual semi classical approximation and the
Renormalization Group makes then us believe that the correct one loop result
is retrieved after integrating the Exact Renormalization Group differential
equation. As a consequence, the perturbative integration of the RG equation
we have proposed $\left( \ref{rgrig}\right) $ seems to be the correct
one-loop wave function renormalization, but contradicts the traditional
perturbative methods.

\section{Conclusion}

We have computed the one-loop quantum correction of the kinetic energy for a
field dependent wave function in one dimension in order to compare various
methods. The main result of the paper is that the perturbative integration
of the ERG equation leads to the exact one loop wave function
renormalization. This result contradicts the traditional perturbative
methods.

\end{document}